\documentclass[11pt,onecolumn,twoside]{IEEEtran} 
\usepackage{graphicx}
\usepackage{amsmath}
\usepackage{mathrsfs}
\usepackage{mathtools}
\usepackage{amssymb}
\usepackage{float}
\usepackage{url}
\usepackage{verbatim}
\usepackage{amsthm}
\usepackage{enumerate}
\usepackage{layout}
\usepackage{bm}

\theoremstyle{definition}

\newcommand{\Reals}{\ensuremath{\mathbb{R}}}
\newcommand{\A}{\ensuremath{\mathcal{A}}}
\newcommand{\expectation}{\ensuremath{\mathbb{E}}}

\begin{document}
\thispagestyle{empty}
\setcounter{page}{1}
\setlength{\baselineskip}{1.15\baselineskip}

\title{\huge{Thermodynamic-kinetic uncertainty relation: properties and an information-theoretic interpretation}\\[0.2cm]}
\author{Tomohiro Nishiyama\\ Email: htam0ybboh@gmail.com}
\date{}
\maketitle
\thispagestyle{empty}

\begin{abstract}
Universal relations that characterize the fluctuations of nonequilibrium systems are of fundamental importance. The thermodynamic and kinetic uncertainty relations impose upper bounds on the precision of currents solely by total entropy production and dynamical activity, respectively. Recently, a tighter bound that imposes on the precision of currents by both total entropy production and dynamical activity has been derived (referred to as the TKUR). In this paper, we show that the TKUR gives the tightest bound of a class of inequalities that imposes an upper bound on the precision of currents by arbitrary functions of the entropy production, dynamical activity, and time interval. Furthermore, we show that the TKUR can be rewritten as an inequality between two Kullback-Leibler divergences. One comes from the ratio of entropy production to dynamical activity, the other comes from the Kullback-Leibler divergence between two probability distributions defined on two-element set, which are characterized by the ratio of precision of the time-integrated current to dynamical activity.
\end{abstract}
 
\section{Introduction}
Studying universal relations that characterize the fluctuations is one of the main themes of nonequilibrium physics. An important class of inequalities with these characteristics is the thermodynamic uncertainty relation (TUR), which imposes an upper bound on the precision of time-integrated currents by entropy production~\cite{barato2015thermodynamic, gingrich2016dissipation, horowitz2017proof, pietzonka2016universal}. On the other hand, the kinetic uncertainty relation (KUR) imposes an another upper bound on generic counting observables by the dynamical activity~\cite{di2018kinetic}. Recently, for discrete-state systems modeled by continuous-time Markov jump processes, a unified bound of the thermodynamic and kinetic uncertainty relation (hereinafter referred to as the TKUR) has been derived~\cite{vo2022unified}. This bound is always tighter than the TUR and KUR, and the bound is written as the product of the dynamical activity and function of the ratio of entropy production to dynamical activity.

In this paper, we give another equivalent forms of the TKUR, and show that the ratio of precision of the time-integrated current to dynamical activity is an important quantity. Next, we consider a class of inequalities that imposes an upper bound on the precision of time-integrated currents by arbitrary functions of the entropy production, dynamical activity, and time interval. This class of inequalities includes the TUR, KUR, TKUR, and extended forms of TUR in Ref.~\cite{horowitz2020thermodynamic}. We show that there does not exist any functions other than the TKUR that always give equal or tighter bounds than the TKUR.  In other words, the TKUR is one of the tightest bounds of this class. Finally, we study the TKUR from the perspective of information theory. The ratio of entropy production to dynamical activity can be interpreted as the Kullback-Leibler divergence (KL-divergence)~\cite{kullback1951information}. We show that the TKUR is rewritten as an inequality between this KL-divergence and the binary KL-divergence, which is the KL-divergence between probability distributions defined on two-element set. The binary KL-divergence is characterized by the ratio of precision of the time-integrated current to dynamical activity. For probability distributions defined on real line, since it is known that the lower bound of KL-divergence under given means and variances is attained by the binary KL-divergence~\cite{nishiyama2020relations, van2020unified, nishiyama2021tight}, the TKUR has a similar property.
 
\section{Preliminaries}
We basically follow the notation of Ref.~\cite{vo2022unified}. We consider a discrete-state system described by a continuous-time Markov jump process. The system is controlled by an arbitrary protocol $\lambda_t=\lambda(vt)$ with speed parameter $v$. The time evolution of the system is given by
\begin{align}
\label{tkur_1_1}
\frac{\mathrm{d}p_n(t,v)}{\mathrm{d}t} = \sum_{m} p_m(t,v)R_{nm}(\lambda_t),
\end{align}
where $p_n(t,v)$ denotes the probability in state $n$ at time $t$ with the speed parameter $v$, and $R_{nm}(\lambda_t)$ denotes the transition rate from state $m$ to state $n$ with the protocol $\lambda_t$. The transition rate satisfies $\sum_{n} R_{nm}(\lambda_t)=0$ and $R_{nm}(\lambda_t)\geq 0$ for $n\neq m$. Let $\omega_\tau=\{n_0, (n_1,t_1), \cdots, (n_N, t_N)\}$ be a trajectory of the system during the time interval $[0,\tau]$. The system is initially at state $n_0$ and jumps from state $n_{i-1}$ to state $n_i$ at time $t_i$ for $1\leq i\leq N$. For each trajectory, let $\mathcal{J}_d(\omega):=\sum_{i=1}^N d_{n_i, n_{i-1}}$ be a generalized time-integrated current, and let coefficients $\{d_{nm}\}$ associated with transition $m\rightarrow n$ be anti-symmetric, $d_{nm}=-d_{mn}$. Assuming that the transition rates satisfy the local detailed balance condition, and letting $K_{nm}(t,v):=p_m(t,v)R_{nm}(\lambda_t)$, the total entropy production $\Sigma_\tau$ and the dynamical activity $\A_\tau$ are given by
\begin{align}
\label{tkur_1_2}
\Sigma_\tau := \int_0^\tau \mathrm{d}t  \sum_{n<m} (K_{nm}(t,v)-K_{mn}(t,v))\ln\frac{K_{nm}(t,v)}{K_{mn}(t,v)}=:\int_0^\tau \mathrm{d}t \sum_{n<m}\sigma_{nm}(t,v),
\end{align}
\begin{align}
\label{tkur_1_3}
\A_\tau := \int_0^\tau \mathrm{d}t  \sum_{n<m} (K_{nm}(t,v)+K_{mn}(t,v))=:\int_0^\tau \mathrm{d}t \sum_{n<m} a_{nm}(v,t),
\end{align}
where $\sigma_{nm}(t,v)$ denotes the entropy production rate, and $a_{nm}(v,t)$ denotes the jump frequency associated with the transition between $m$ and $n$.
Next, for arbitary function $F$, we consider a class of inequalities such that
\begin{align}
\label{tkur_1_4}
F(\Sigma_\tau, \A_\tau, \tau) \geq \frac{(\nabla \expectation[\mathcal{J}_d])^2}{\mathrm{Var}[\mathcal{J}_d]}.
\end{align}
Here, $\nabla:=\tau\partial_\tau - v\partial_v$ is a differential operator, and $\expectation[\cdot]$ and $\mathrm{Var}[\cdot]$ denote the ensemble average and variance of the current.
The TUR takes $F(\Sigma_\tau, \A_\tau, \tau)=\frac{\Sigma_\tau}{2}$, and the KUR takes $F(\Sigma_\tau, \A_\tau, \tau)=\A_\tau$~\cite{koyuk2020thermodynamic}. The TUR and the KUR connects the total entropy production and the dynamical activity with the precision of the time-integrated current, respectively.
Furthermore, the TKUR takes
\begin{align}
\label{tkur_1_5}
 F(\Sigma_\tau, \A_\tau, \tau)=\frac{\A_\tau S_\tau^2}{4 f\Bigl(\frac{S_\tau}{2}\Bigr)^2},
\end{align}
where $S_\tau:=\Sigma_\tau \A_\tau^{-1}$ and $f(\cdot)$ denotes the inverse function of $x\mathrm{tanh}(x)$.
The TKUR is always tighter than the TUR and KUR. Finally, we define the pseudo-entropy production, which is an empirical measure of irreversibility, as follows.
\begin{align}
\label{tkur_1_6}
\Sigma_\tau^{\mathrm{ps}} := 2\int_0^\tau \sum_{n<m} \frac{j_{nm}(t,v)^2}{a_{nm}(t,v)},
\end{align}
where $j_{nm}(t,v):=K_{nm}(t,v)-K_{mn}(t,v)$ is the probability current from state $m$ to state $n$.
The pseudo-entropy production satisfies the following inequality (see Ref.~\cite{vo2022unified}).
\begin{align}
\label{tkur_1_7}
\Sigma_\tau^{\mathrm{ps}} \leq \frac{\A_\tau S_\tau^2}{2 f\Bigl(\frac{S_\tau}{2}\Bigr)^2}.
\end{align}

\section{Another forms of the TKUR}
We derive another equivalent forms of the TKUR and  a time-integrated form of the TKUR in the short-time limit. Let $g(\cdot)$ be the inverse function of $x\mathrm{artanh}(x)$ for non-negative $x$, where $\mathrm{artanh}(x)=\frac12 \ln\frac{(1+x)}{(1-x)}$. For $x\geq 0$, we prove
\begin{align}
\label{tkur_2_1}
\frac{x}{f(x)}=g(x).
\end{align} 
\begin{proof}
Substituting $x=\mathrm{tanh}(y)$ into $g(x\mathrm{artanh}(x))=x$, we have 
\begin{align}
\label{tkur_2_2}
g\Bigl(y\mathrm{tanh}(y)\Bigr)=\mathrm{tanh}(y).
\end{align}
On the other hand, substituting $x=y\mathrm{tanh}(y)$ into $\frac{x}{f(x)}$, we have
\begin{align}
\label{tkur_2_3}
\frac{y\mathrm{tanh}(y)}{f(y\mathrm{tanh}(y))}=\mathrm{tanh}(y),
\end{align}
where we use the definition of $f(y\mathrm{tanh}(y))=y$.
Since $y\mathrm{tanh}(y)$ is an arbitrary positive real number, from \eqref{tkur_2_2} and \eqref{tkur_2_3}, we obtain \eqref{tkur_2_1}.
\end{proof}
From \eqref{tkur_2_1}, the TKUR can be rewritten as 
\begin{align}
\label{tkur_2_4}
\A_\tau g\Bigl(\frac{S_\tau}{2}\Bigr)^{2}\geq \frac{\Bigl(\nabla \expectation[\mathcal{J}_d]\Bigr)^2}{\mathrm{Var}[\mathcal{J}_d]}.
\end{align}
Since the function $g(\cdot)$ is monotonically increasing, we also obtain
\begin{align}
\label{tkur_2_5}
S_\tau \geq  2\mathcal{R}\mathrm{artanh}(\mathcal{R})=\mathcal{R}\ln\frac{1+\mathcal{R}}{1-\mathcal{R}},
\end{align}
where 
\begin{align}
\label{tkur_2_6}
\mathcal{R}:=\frac{|\nabla \expectation[\mathcal{J}_d]|}{\sqrt{\A_\tau \mathrm{Var}[\mathcal{J}_d]}}.
\end{align}
The quantity $\mathcal{R}$ is equal to the square root of the ratio of both sides of the KUR. From \eqref{tkur_2_5}, the lower bound of entropy production (to be exact $S_\tau$) increases as $\mathcal{R}$ increases, and the entropy production diverges when $\mathcal{R}$ goes to $1$. Hence, $\mathcal{R}$ can be interpreted as the measure of irreversibility. We show an example of  $\mathcal{R}=1$ in Appendix~\ref{ap_1}. When $\mathcal{R}$ approaches zero, the inequality \eqref{tkur_2_5} reduces to the TUR. From \eqref{tkur_2_4} and  \eqref{tkur_2_5}, we can derive the KUR and TUR immediately since the range of the function $g$ is in $[0,1]$, and $\mathrm{artanh}(x)\geq x$.
In the short time limit $\tau\rightarrow 0$, the inequality \eqref{tkur_2_4} yields
\begin{align}
\label{tkur_2_7}
g\Bigl(\frac{\sum_{n<m}\sigma_{nm}(t,v)}{2\sum_{n<m}a_{nm}(t,v)}\Bigr)^{2} \sum_{n<m}a_{nm}(t,v)\geq \frac{\Bigl(\sum_{n<m}d_{nm}j_{nm}(t,v)\Bigr)^2}{\sum_{n<m}d_{nm}^2a_{nm}(t,v)}.
\end{align}
From the Cauchy-Schwartz inequality, we have
\begin{align}
\label{tkur_2_8}
\int_0^\tau \mathrm{d}t \sum_{n<m} \frac{\Bigl(K_{nm}(t,v)-K_{mn}(t,v)\Bigr)^2}{K_{nm}(t,v)+K_{mn}(t,v)}\int_0^\tau\mathrm{d}t \sum_{n<m} d_{nm}^2(K_{nm}(t,v)+K_{mn}(t,v)) \nonumber \\
\geq \Bigl(\int_0^\tau \mathrm{d}t \sum_{n<m} d_{nm}(K_{nm}(t,v)-K_{mn}(t,v))\Bigr)^2.
\end{align}
From this relation and \eqref{tkur_1_6}, we have
\begin{align}
\label{tkur_2_9}
\frac{\Sigma_\tau^{\mathrm{ps}}}{2}\geq \frac{\expectation[\mathcal{J}_d]^2}{\expectation[\mathcal{J}_{d^2}]},
\end{align}
where $\expectation[\mathcal{J}_{d^2}]$ denotes the ensemble average of the symmetric current for $d_{nm}^2=d_{mn}^2$.
By combining this relation with \eqref{tkur_1_7} and \eqref{tkur_2_1}, we have 
\begin{align}
\label{tkur_2_10}
\A_\tau g\Bigl(\frac{S_\tau}{2}\Bigr)^{2}\geq \frac{\expectation[\mathcal{J}_d]^2}{\expectation[\mathcal{J}_{d^2}]}.
\end{align}
This inequality is a time-integrated form of \eqref{tkur_2_7}, and it is tighter than time-integrated form of the short-time TUR~\cite{otsubo2020estimating}.
We also have the same inequality as \eqref{tkur_2_5} for 
\begin{align}
\label{tkur_2_11}
\tilde{\mathcal{R}}:=\frac{|\expectation[\mathcal{J}_d]|}{\sqrt{\A_\tau\expectation[\mathcal{J}_{d^2}]}}.
\end{align}

\section{The tightest bound of the class of inequalities \eqref{tkur_1_4}}
\label{sec_tightest}
In this section, for the class of inequalities \eqref{tkur_1_4}, we prove $F(\Sigma_\tau, \A_\tau, \tau)=\A_\tau g\Bigl(\frac{S_\tau}{2}\Bigr)^2 $ if $F(\cdot)$ satisfies
\begin{align}
\label{tkur_3_2}
\A_\tau g\Bigl(\frac{S_\tau}{2}\Bigr)^2\geq F(\Sigma_\tau, \A_\tau, \tau) \geq \frac{\Bigl(\nabla \expectation[\mathcal{J}_d]\Bigr)^2}{\mathrm{Var}[\mathcal{J}_d]}.
\end{align}
This result means that there does not exist any functions $F(\cdot)$ other than the TKUR that always give equal or tighter bounds than the TKUR. 
\begin{proof}
We consider a steady-state such that 
\begin{align}
\label{tkur_3_3}
\A_\tau&=\tau(\alpha+\beta), \\
\label{tkur_3_4}
\Sigma_\tau&=\tau(\alpha-\beta)\ln\frac{\alpha}{\beta}, \\
\label{tkur_3_5}
\expectation[\mathcal{J}_d]&=\tau d(\alpha-\beta), \\
\label{tkur_3_6}
\mathrm{Var}[\mathcal{J}_d]&=\tau d^2(\alpha+\beta),
\end{align} 
where $d:=d_{nm}=-d_{mn}$ and $\alpha\neq\beta$.
In Appendix \ref{ap_1}, we show an example that satisfies \eqref{tkur_3_3}-\eqref{tkur_3_6}.
Without any loss of generality, we can define a function $G(x,y,z)$ as 
\begin{align}
\label{tkur_3_7}
G\Bigl(\frac{\Sigma_\tau}{\A_\tau}, \frac{\A_\tau}{\tau}, \tau\Bigr):=F(\Sigma_\tau, \A_\tau, \tau).
\end{align}
Substituting \eqref{tkur_3_3}-\eqref{tkur_3_6} and \eqref{tkur_3_7} into \eqref{tkur_3_2}, and dividing by $\A_\tau$, we have
\begin{align}
\label{tkur_3_8}
g\Bigl(\frac{\gamma}{2}\Bigr)^2\geq \frac{1}{\tau(\alpha+\beta)}G(\gamma, \alpha+\beta, \tau) \geq \frac{(\alpha-\beta)^2}{(\alpha+\beta)^2}.
\end{align}
where $\gamma:=\frac{\alpha-\beta}{\alpha+\beta}\ln\frac{\alpha}{\beta}$.
Substituting $x=\frac{\alpha-\beta}{\alpha+\beta}$ into $g(x\mathrm{artanh}x)^2=x^2$, we have
\begin{align}
\label{tkur_3_9}
g\Bigl(\frac{\gamma}{2}\Bigr)^2=\frac{(\alpha-\beta)^2}{(\alpha+\beta)^2}.
\end{align}
From \eqref{tkur_3_8} and \eqref{tkur_3_9}, it follows that
\begin{align}
\label{tkur_3_10}
\frac{(\alpha-\beta)^2}{(\alpha+\beta)^2}=\frac{1}{\tau(\alpha+\beta)}G(\gamma, \alpha+\beta, \tau).
\end{align}
Since the left hand-side does not depend on $\tau$, we can write
\begin{align}
\label{tkur_3_11}
\frac{1}{\tau(\alpha+\beta)}G(\gamma, \alpha+\beta, \tau)=\frac{1}{\alpha+\beta}\tilde{G}(\gamma, \alpha+\beta).
\end{align}
Furthermore, by defining $X:=\frac{\beta}{\alpha}$, we have 
\begin{align}
\label{tkur_3_12}
\frac{(1-X)^2}{(1+X)^2}=\frac{1}{\alpha(1+X)}\tilde{G}(\gamma[X], \alpha(1+X)),
\end{align}
where we write $\gamma[X]=\frac{X-1}{X+1}\ln X$ to emphasize that $\gamma$ depends only on $X$.
Since the left hand-side of \eqref{tkur_3_12} does not depend on $\alpha$, we can write
\begin{align}
\label{tkur_3_13}
\hat{G}(\gamma[X]):=\frac{1}{\alpha(1+X)}\tilde{G}(\gamma[X], \alpha(1+X))
\end{align}
Since $\gamma[X]$ takes an arbitrary non-negative real number, by comparing $\eqref{tkur_3_9}$ with \eqref{tkur_3_12} and \eqref{tkur_3_13}, we have
$\hat{G}(x)=g(\frac{x}{2})^2$ for $x\geq 0$.
By combining this relation with \eqref{tkur_3_7}, \eqref{tkur_3_11}, and \eqref{tkur_3_13}, we have $F(\Sigma_\tau, \A_\tau, \tau)=G\Bigl(S_\tau, \frac{\A_\tau}{\tau}, \tau\Bigr)=\tau\tilde{G}\Bigl(S_\tau, \frac{\A_\tau}{\tau}\Bigr)=\A_\tau\hat{G}\Bigl(S_\tau\Bigr)=\A_\tau g\Bigl(\frac{S_\tau}{2}\Bigr)^2$.
Hence, the result follows.
\end{proof}
The same discussion can be applied for 
\begin{align}
\label{tkur_3_14}
F(\Sigma_\tau, \A_\tau, \tau) \geq \frac{\expectation[\mathcal{J}_d]^2}{\expectation[\mathcal{J}_{d^2}]}.
\end{align}
In this case, the inequality \eqref{tkur_2_10} gives one of the tightest bounds of inequalities \eqref{tkur_3_14}.  In \eqref{tkur_1_4}, given the class of inequalities with $\mathcal{J}_d$ replaced by a generic counting observables $O$, the KUR $F(\Sigma_\tau, \A_\tau, \tau)=\A_\tau$ gives one of the tightest bounds of this class (see Appendix~\ref{ap_3}).

\section{Information-theoretic interpretation of the TKUR}
In this section, we show that the both sides of \eqref{tkur_2_5} can be interpreted as the KL-divergence. The KL-divergence between probability distributions $P$ and $Q$ on the same probability space $\chi$ is defined by
\begin{align}
\label{tkur_4_1}
D(P\|Q):=\sum_{\chi} P(\chi)\ln \frac{P(\chi)}{Q(\chi)}.
\end{align}
Let $\mathcal{S}$ be a set of states, and let $P$ and $Q$ be probability distributions defined on the set $\{(n, m, t) | n,m \in \mathcal{S}, n\neq m, t \in [0, \tau]\}$ as follows.
\begin{align}
\label{tkur_4_2}
&P(n,m,t):=\frac{1}{\A_\tau} K_{nm}(t,v), \\
\label{tkur_4_3}
&Q(n,m,t):=\frac{1}{\A_\tau}K_{mn}(t,v) =P(m,n,t), \\
&\int_0^\tau\mathrm{d}t\sum_{n\neq m} P(n,m,t)=1,
\end{align}
where we drop the parameter $v$ of the left-hand side. 
From \eqref{tkur_1_2} and the definition of the KL-divergence \eqref{tkur_4_1}, we have 
\begin{align}
\label{tkur_4_4}
S_\tau=D(P\|Q).
\end{align}
For $\phi_1, \phi_2\in\Reals$, let $P_2$ and $Q_2$ be probability distributions defined on two-element set $\{\phi_1, \phi_2\} $ as follows. 
\begin{align}
\label{tkur_4_4}
P_2(\phi_1)=\frac{1+\mathcal{R}}{2}, \quad P_2(\phi_2)=\frac{1-\mathcal{R}}{2}, \\
\label{tkur_4_5}
Q_2(\phi_1)=\frac{1-\mathcal{R}}{2}, \quad Q_2(\phi_2)=\frac{1+\mathcal{R}}{2}.
\end{align}
Since $\mathcal{R}$ is equal to the square root of the ratio of both sides of the KUR, the KUR $\mathcal{R}^2 \leq 1$ guarantees that $P_2$ and $Q_2$ are probability distributions. 
From \eqref{tkur_2_5}, we have
\begin{align}
\label{tkur_4_6}
S_\tau=D(P\|Q)\geq D(P_2\|Q_2),
\end{align}
where we use $D(P_2\|Q_2)=\mathcal{R}\ln\frac{(1+\mathcal{R})}{(1-\mathcal{R})}$. From \eqref{tkur_4_6}, the KL-divergence between $P$ and $Q$ under given quantity $\mathcal{R}$ is lower bounded by the binary KL-divergence $D(P_2\|Q_2)$. Since the lower bound of KL-divergence between probability distributions defined on real line $\Reals$ under given means and variances is attained by the binary divergence~(see \cite{nishiyama2020relations}), the inequality \eqref{tkur_4_6} has the similar property.
By defining $\tilde{P}_2$ and $\tilde{Q}_2$ for $\tilde{\mathcal{R}}$ in \eqref{tkur_2_11} in the similar way to \eqref{tkur_4_4} and \eqref{tkur_4_5}, from \eqref{tkur_2_5} in the case of $\tilde{\mathcal{R}}$, we obtain
\begin{align}
\label{tkur_4_7}
D(P\|Q)\geq D(\tilde{P}_2\|\tilde{Q}_2).
\end{align}
Letting $\phi(n,m,t):=d_{nm}$ be a random variable, the means and variances of $\phi(n,m,t)$ with respect to $P$ and $Q$ are given by
\begin{align}
\label{tkur_4_8}
&\expectation[\phi]_{P}= -\expectation[\phi]_{Q}=\int_0^\tau\mathrm{d}t\sum_{n\neq m} \phi(n,m,t) P(n,m,t)=\frac{\expectation[\mathcal{J}_d]}{\A_\tau}, \\
\label{tkur_4_9}
&\mathrm{Var}[\phi]_{P}=\mathrm{Var}[\phi]_{Q}=\int_0^\tau\mathrm{d}t\sum_{n\neq m} \phi(n,m,t)^2 P(n,m,t)-\expectation[\phi]_{P}^2=\frac{\expectation[\mathcal{J}_{d^2}]}{\A_\tau}-\Bigl(\frac{\expectation[\mathcal{J}_d]}{\A_\tau}\Bigr)^2.
\end{align} 
By choosing $\{\phi_1, \phi_2\}=\{\mathrm{sign}(\expectation[\mathcal{J}_d])\sqrt{\frac{\expectation[\mathcal{J}_{d^2}]}{\A_\tau}}, -\mathrm{sign}(\expectation[\mathcal{J}_d])\sqrt{\frac{\expectation[\mathcal{J}_{d^2}]}{\A_\tau}}\}$, it can be verified that the means and variances of a random variable  $\phi$ with respect to $\tilde{P}_2$ and $\tilde{Q}_2$ are equal to \eqref{tkur_4_8} and \eqref{tkur_4_9}, respectively. Here, $\mathrm{sign}(\cdot)$ is the sign function such that $\mathrm{sign}(x):=1 \; \mbox{for} \; x \geq 0$, and $\mathrm{sign}(x):=-1 \; \mbox{for} \; x < 0$. In Appendix \ref{ap_2}, we prove the inequality \eqref{tkur_4_7} by a different approach.
\section{Conclusion} 
In this paper, we have provided another forms of the TKUR. Considering the class of inequalities that imposes an upper bound on the precision of time-integrated currents by arbitrary functions of the entropy production, dynamical activity, and time interval, we have shown that the TKUR gives one of the tightest bounds of this class. In other words, there does not exist any functions other than the TKUR that always give equal or tighter bounds than the TKUR. Furthermore, we have interpreted the ratio of entropy production to dynamical activity as the KL-divergence, and we have shown that it is lower bounded by the binary KL-divergence. This property is similar to the lower bound of the KL-divergence between probability distributions defined on real line under given means and variances.

\appendices
\section{Example of the system that satisfies \eqref{tkur_3_3}-\eqref{tkur_3_6}}
\label{ap_1} 
We consider a three-state Markov jump process with fully connected states and time-independent transition rates. Let $R_{32}=R_{21}=R_{13}=:\alpha$, $R_{12}=R_{23}=R_{31}=:\beta$, and let $p_1(t=0)=p_2(t=0)=p_3(t=0)=\frac13$. Clockwise and counterclockwise have different transition rate $\alpha\geq 0$ and $\beta\geq 0$, respectively. For the same direction that the transition rate is $\alpha$, let $d_{32}=d_{21}=d_{13}=:d$. From these assumptions, this system is in steady-state, and it can be easily verified that \eqref{tkur_3_3}-\eqref{tkur_3_5} hold. Regarding \eqref{tkur_3_6}, let $\{X_i\}$ be random variables defined on $\{-d, 0, d\}$ with probabilities $\{\beta \mathrm{d}t, 1-(\alpha+\beta)\mathrm{d}t, \alpha\mathrm{d}t\}$ for $1\leq i \leq N$.
The variance of the time-integrated current is given by $\sum_{k=1}^3 p_k \mathrm{Var}[\sum_{i=1}^N X_i]=\sum_{i=1}^N \mathrm{Var}[X_i]=\tau d^2(\alpha+\beta)$, where we use $X_i$ and $X_j$ are independent for $i\neq j$, and we neglect $O(\mathrm{d}t)$.
From \eqref{tkur_2_6} and \eqref{tkur_3_3}-\eqref{tkur_3_6}, it can be verified that $\mathcal{R}=1$, and the entropy production diverges when $\alpha=0$ or $\beta=0$. In this case, only clockwise or counterclockwise transitions can occur.

\section{Another proof of \eqref{tkur_4_7}}
\label{ap_2}
Let $\tilde{P}$ and $\tilde{Q}$ be probability distributions of a random variable $\phi$ as follows.
\begin{align}
\label{ap2_1}
\tilde{P}(\phi)&:=\int_0^\tau\mathrm{d}t\sum_{n\neq m} \delta\Bigl(\phi, \phi(n,m,t)\Bigr) P(n,m,t), \\
\label{ap2_2}
\tilde{Q}(\phi)&:=\int_0^\tau\mathrm{d}t\sum_{n\neq m} \delta\Bigl(\phi, \phi(n,m,t)\Bigr) Q(n,m,t)=\tilde{P}(-\phi),
\end{align}
where $\delta(x,y)$ denotes the Kronecker delta and we use \eqref{tkur_4_3}.
These probability distributions are defined on $\Reals$, and the means and variances of $\phi$ with respect to $\tilde{P}$ and $\tilde{Q}$ are given by
\begin{align}
\label{ap2_3}
\expectation[\phi]_{\tilde{P}}  = -\expectation[\phi]_{\tilde{Q}}  &=\int_0^\tau\mathrm{d}t\sum_{n\neq m} \phi(n,m,t) P(n,m,t)=\expectation[\phi]_{P}, \\
\label{ap2_4}
\mathrm{Var}[\phi]_{\tilde{P}}=\mathrm{Var}[\phi]_{\tilde{Q}}  &=\int_0^\tau\mathrm{d}t\sum_{n\neq m} \phi(n,m,t)^2 P(n,m,t)-\expectation[\phi]_{\tilde{P}}^2=\mathrm{Var}[\phi]_P,
\end{align}
where we use \eqref{tkur_4_8} and \eqref{tkur_4_9}.
From the chain rule for the KL-divergence~\cite{cover1999elements}, we have
\begin{align}
\label{ap2_5}
D(P\|Q)\geq D(\tilde{P}\|\tilde{Q}).
\end{align}
From Theorem 2 in Ref.~\cite{nishiyama2020relations}, the lower bound of $D(\tilde{P}\|\tilde{Q})$ under conditions \eqref{ap2_3} and \eqref{ap2_4} is attained by the binary KL-divergence between probability distributions $\tilde{P}_2$ and $\tilde{Q}_2$ defined on $\{\phi_1, \phi_2\}$ such that
\begin{align}
\label{ap2_6}
&\tilde{P}_2(\phi_1)=\frac{1+\tilde{\mathcal{R}}}{2}, \quad \tilde{Q}_2(\phi_1)=\frac{1-\tilde{\mathcal{R}}}{2}, \\
&\phi_1=\mathrm{sign}(\expectation[\mathcal{J}_d])\sqrt{\frac{\expectation[\mathcal{J}_{d^2}]}{\A_\tau}}, \quad \phi_2=-\mathrm{sign}(\expectation[\mathcal{J}_d])\sqrt{\frac{\expectation[\mathcal{J}_{d^2}]}{\A_\tau}}, \\
&\expectation[\phi]_{\tilde{P}_2}=\expectation[\phi]_{\tilde{P}}, \quad \mathrm{Var}[\phi]_{\tilde{P}_2}=\mathrm{Var}[\phi]_{\tilde{P}}, \\
&\expectation[\phi]_{\tilde{Q}_2}=\expectation[\phi]_{\tilde{Q}}, \quad \mathrm{Var}[\phi]_{\tilde{Q}_2}=\mathrm{Var}[\phi]_{\tilde{Q}}.
\end{align}
The relation \eqref{ap2_6} corresponds to \eqref{tkur_4_4} and \eqref{tkur_4_5} for $\tilde{\mathcal{R}}$. By combining $D(\tilde{P}\|\tilde{Q})\geq D(\tilde{P}_2\|\tilde{Q}_2)$ with \eqref{ap2_5}, we obtain \eqref{tkur_4_7}.

\section{Tightness of the KUR}
We use the same notation as Section~\ref{sec_tightest}.
\label{ap_3}
We show that $F(\Sigma_\tau, \A_\tau, \tau)=\A_\tau$ if there exists $F(\cdot)$ such that
\begin{align}
\label{ap3_1}
\A_\tau \geq F(\Sigma_\tau, \A_\tau, \tau) \geq \frac{\Bigl(\nabla \expectation[O]\Bigr)^2}{\mathrm{Var}[O]},
\end{align}
where $O$ denotes the generic counting observables. As $O$, choosing a symmetric current $\mathcal{J}_d$ such that $d_{nm}=d_{mn}$ for $n\neq m$ and $d_{mm}=0$, and considering the system in Appendix~\ref{ap_1}, we obtain \eqref{tkur_3_3}, \eqref{tkur_3_4}, \eqref{tkur_3_6}, and $\expectation[\mathcal{J}_d]=\tau d(\alpha+\beta)$ instead of \eqref{tkur_3_5}. Substituting these relations into \eqref{ap3_1}, we have
\begin{align}
\label{ap3_2}
1=\frac{1}{\tau(\alpha+\beta)}G(\gamma, \alpha+\beta, \tau).
\end{align}
In the similar way to Section~\ref{sec_tightest}, we have $\hat{G}(x)=1$.
Hence, we obtain  $F(\Sigma_\tau, \A_\tau, \tau)=G\Bigl(S_\tau, \frac{\A_\tau}{\tau}, \tau\Bigr)=\tau\tilde{G}\Bigl(S_\tau, \frac{\A_\tau}{\tau}\Bigr)=\A_\tau\hat{G}\Bigl(S_\tau\Bigr)=\A_\tau$.

\bibliography{reference_TUR} 
\bibliographystyle{myplain} 
\end{document}